\documentclass[fleqn,11pt,a4paper]{article}

\usepackage{epsfig,amsmath,amssymb,bm}
\usepackage{color}

\def\be{\begin{equation}}
\def\ee{\end{equation}}
\def\bea{\begin{eqnarray}}
\def\eea{\end{eqnarray}}
\def\ba{\begin{array}}
\def\ea{\end{array}}
\def\bem{\begin{displaymath}}
\def\enm{\end{displaymath}}
\def\d{\textrm{d}}
\def\c{\chi}

\def\w{\wedge}
\def\wt{\bar}

\def\h{\hat}

\def\nn{\nonumber}

\def\d{\textrm{d}}
\def\c{\chi}
\def\D{\textrm{D}}
\def\i{\textrm{i}}

\def\R{\textrm{R}}

\begin{document}
\title{\bfseries Charged Fluid Dynamics in Scalar-Tensor Theories of Gravity with Torsion}
\author{\bfseries Chih-Hung Wang\thanks{Institute of Physics, Academia Sinica, Taipei 115, Taiwan, Republic of China(email : chhwang@phys.sinica.edu.tw)} \thanks{Department of Physics,
Lancaster University, UK (email : robbin1101@hotmail.com)}}

\maketitle 

\begin{abstract}

In scalar-tensor theories of gravity with torsion, the
gravitational field is described in terms of a symmetric metric
tensor $g$, a metric-compatible connection $\nabla$ with torsion,
and a scalar field $\varphi$. The main aim is to explore an
interaction of a charged perfect fluid and a scalar field
$\varphi$ in a background electromagnetic and gravitational field
described by \{$g$, $\nabla$, $\varphi$\}. The interaction is
based on an action functional $S_C$ of a charged perfect fluid
that is invariant under global conformal rescalings. Using a
variational principle, we obtain equations of motion for the
charged perfect fluid. Moreover, we verify that these equations of
motion are equivalent to the gauge identities obtained from the
invariance of an action functional under spacetime dffeomorphisms
and a local U(1) gauge group.
\end{abstract}

\section{Action Functionals of Charged Perfect Fluids \label{action}}

In Brans-Dicke theory \cite{BD61} a direct interaction between the
Brans-Dicke scalar field and matter is often said to violate the
weak equivalence principle \cite{Will06}. However, as Dirac
\cite{Dirac73} has shown one may consider gravity in a Weyl
geometry which naturally induces such interactions. But little
exploration has been made of gravitational scalar field
interactions in a Cartan geometry. Dereli and Tucker \cite{DT02}
have noted that the motion of an electrically neutral spinless
particle in the background geometry of Dirac's theory can also be
reformulated in terms of an autoparallel of a Cartan geometry with
torsion determined by the gradient of the Brans-Dicke scalar field
\cite{DT82}. In this note, we examine the motion of a relativistic
charged fluid
in a background Cartan
geometry determined by a scalar field. We shall motivate a
particular coupling of such a field to the fluid and derive the
equations of motion from an action principle.

In general relativity (GR), the theory of a relativistic perfect
fluid with internal energy $\varepsilon(n)$, number density $n$,
4-velocity field $V$ composed of particles of rest mass $m$ and
electric charge $q$ can be determined from the action
%
\bea
{S}_{GR}& =& \int_M\Lambda_{GR}(e^a, A; V, n,\,\lambda_1, \lambda_2) \nn\\
&=&\int_M\left(- m\,n(1+\varepsilon(n))\star1 + q\,n \,A\w \star
\widetilde{V} + \lambda_1\,
\d\,(n\star\widetilde{V})+\lambda_2\,(\widetilde{V}\w\star\widetilde{V}+
\star 1)\right), \nn
\eea where the background metric of spacetime is $g = e^a \otimes
e_a$ and the background Maxwell field is $F=\d A$. Latin indices
run over 0 to 3. The signature is $(-, +, +, +)$ and $c=1$. The
Lagrange multiplier fields $\lambda_1$, $\lambda_2$ ensure the
conservation equation $\d\,(n\star\widetilde{V})=0$ and
normalization of 4-velocity $g(V,V)=-1$. We use $;$ in
$\Lambda_{GR}$ to separate background fields $e^a$, $A$ from the
fluid's variables.
$\widetilde{V}$ denotes the metric dual of $V$, i.e.
$\widetilde{V}\equiv g(V, -)$, and $\star$ is the Hodge map. For
any 1-form $\alpha$, $\widetilde{\alpha}= g^{-1}(\alpha, -)$,
where $g^{-1}$ is the inverse of $g$.
In a non-Riemannian Cartan background with specified metric $g$
and torsion forms $T^a$, we suppose that the gravitational scalar
$\varphi$ couples to the relativistic fluid to modify the action
to \bea
{S}_{C} & =& \int_M\Lambda_{C}(e^a, A, \varphi\,; V, n,\,\lambda_1,\lambda_2)\nn\\
&=&\int_M\left(-\c\,\varphi\,n\,(1+\varepsilon(n))\star1 +
q\,n \,A\w \star \widetilde{V} + \lambda_1\,
\d\,(n\star\widetilde{V})+\lambda_2\,(\widetilde{V}\w\star\widetilde{V}+
\star 1)\right),\nn
\eea where $\c$ is a coupling constant. We have motivated the
action by demanding that $S_C|_{\varphi=\frac{m}{\c}}=S_{GR}$ and
that it be invariant under the following global rescalings
\bea
&&\wt{e}^a = \Omega\, e^a, \nn \hspace{1.0cm} \wt{A} = A, \nn
\hspace{1.0cm} \wt{\varphi} = \Omega^{-1}\, \varphi, \nn \\
&&\wt{V} = \Omega^{-1}\, V, \nn \hspace{0.5cm} \wt{n} =
\Omega^{-3}\,n, \hspace{0.5cm} \wt{\varepsilon}(n)= \varepsilon,
\hspace{0.5cm} \wt{\lambda}_1 = \lambda_1, \hspace{0.5cm}
\wt{\lambda}_2= \Omega^{-4} \lambda_2, \nn
\eea where $\Omega=\textrm{constant}$. Here, the conformal weights
of $V$ and $n$ are obtained by requiring that
$\d\,(n\star\widetilde{V})=0$ and $g(V,V)=-1$ remain invariant
under $\Omega$ transformations.
Furthermore, we assume that $\c$ and $q$ are also invariant under
$\Omega$ transformations.

\section{Equations of Motion for the Charged Perfect Fluid}

By varying $S_C$ with respect to $V$, $n$, $\lambda_1$, and
$\lambda_2$, we obtain
\bea
&&q\,n\star A-n\star \d\,\lambda_1 +
2\lambda_2\star\widetilde{V}=0,
\label{varyVphi}\\
&&-\c\,\varphi \,f\,\star 1
+ q\,A\w\star\widetilde{V} - \d\lambda_1\w\star\widetilde{V}=0, \label{varyNphi}\\
&&\d\,(n\star\widetilde{V})=0, \label{varylambda1phi}\\
&&\widetilde{V}\w\star\widetilde{V}+\star 1=0,
\label{varylambda2phi}
\eea where $f=1 + \varepsilon(n) + n\,\frac{\d\varepsilon}{\d n}$
is the index of the fluid \cite{Lich67}. By solving for
$\lambda_1$ and $\lambda_2$, we can derive the equations of motion
for $V$ and $n$ by substituting them into the rest of equations.
There is no obvious way to know which two equations are best used
for solving $\lambda_1$ and $\lambda_2$ since they both appear in
the 3-form equation (\ref{varyVphi}). Using (\ref{varyVphi}),
(\ref{varyNphi}) and (\ref{varylambda2phi}), we can first solve
for $\lambda_2$ to obtain
$\lambda_2=\frac{1}{2}\c\,\varphi\, n\,f.$
%
Then, we eliminate $\lambda_1$ from (\ref{varyVphi}) and take the
interior derivative $\i_V$ to get
\bea
q\,\i_V\,F+\i_V\,\d(\c\,\varphi\,f\widetilde{V})=0,
\label{elimatedlambda_1_2}
\eea where the solution of $\lambda_2$ has been used and $F=\d A$.
Because of the identity $\i_V\,\i_V \theta\equiv0$ for any p-form
$\theta$, one recognizes that the 1-form equation
(\ref{elimatedlambda_1_2}) has only 3 independent scalar
equations. We consider the 3 independent equations
(\ref{elimatedlambda_1_2}) with (\ref{varylambda1phi}) and
(\ref{varylambda2phi}) as equations of motion for $V$ and $n$. It
is interesting to notice that equations
(\ref{elimatedlambda_1_2}), (\ref{varylambda1phi}) and
(\ref{varylambda2phi}) do not involve any direct connection
couplings. We can express (\ref{elimatedlambda_1_2}) in terms of
the torsion free Levi-Civita connection $\h{\nabla}$ as \bea
\,\c\,\varphi\,n\,f\,\hat{\nabla}_V V + q\,n\,\widetilde{\i_V F} +
\c\,\varphi\,\Pi_V\widetilde{\hat{\nabla}\, {p}} +\c\,n\,f\,
\Pi_V\widetilde{\hat{\nabla}\,\varphi} =0, \label{T=01}
\eea where $p = n^2\frac{\d \varepsilon}{\d n} $ is the pressure
per mass and $\Pi_V$ is a projection operator defined by
$\Pi_V\equiv \textbf{1} + \, V \otimes \widetilde{V}$. From
(\ref{T=01}), we find that $\varphi$ has a similar effect as the
pressure per mass $p$ acting on the motion of the fluid. Moreover,
a neutral dust-like ($q=p=0$) fluid does not follow geodesic flow
due to the scalar field effect. If we consider the background
torsion field $T^a= e^a \w \frac{\d \varphi}{\varphi}
$\cite{DT82}, (\ref{T=01}) becomes
\bea
\,n\,\c\,\varphi\,f\,\nabla_V V + q\,n\,\widetilde{\i_V F} +
\c\,\varphi\,\Pi_V\widetilde{\hat{\nabla}\,{p}}=0,
\label{torsionvarphi2}
\eea where the non-Riemannian connection involves
$\varphi$.
From (\ref{torsionvarphi2}), it follows that the integral curves
of $V$ for a neutral dust-like fluid are autoparallels of $\nabla$
rather than $\hat{\nabla}$.

Invariance of $S_{C}$ under spacetime diffeomorphism, local
SO(3,1), and U(1) gauge transformations yield the gauge identities
\cite{wang06}
\bea
&&\D\tau_a + \tau_c \w \i_{X_a}T^c + S_b{^c} \w \i_{X_a} \R^b{_c}
+ j \w \i_{X_a}F - \rho\,(\i_{X_a}\d\varphi) \equiv 0,
\label{gaugeidentity2}\\
&&\D\,S_a{^b}-\frac{1}{2}(\tau_a\w e^b-\tau^b\w e_a)\equiv 0,
\label{gaugeidentity4}\\
&&\d j \equiv0, \label{gaugeidentity5}
\eea where 3-forms $\tau_a$, $S_a{^b}$, $j$, and 4-form $\rho$ can
be recognized as source currents of $e^a$, connection 1-forms
$\omega^a{_b}$ with torsion, $F$, and $\varphi$ respectively. $\D$
is a covariant exterior derivative with respect to $\omega^a{_b}$
and $\R^b{_c}$ are the associated curvature 2-forms. From $S_C$,
these source currents become \bea
\tau_a&=&\c\,\varphi\,(\,V_a(\,n(\,1 + \varepsilon\,) +
{p})\star\widetilde{V} + {p}\star e_a\,), \nn\\
j&=&q\,n\,\star V, \nn\\
{\rho}&=&-\c\,n(\,1 + \varepsilon\,)\star1, \nn
\eea and $S_a{^b}=0$. \cite{wang06} By substituting these source
currents into (\ref{gaugeidentity2}), (\ref{gaugeidentity4}) and
(\ref{gaugeidentity5}), we find that (\ref{varylambda1phi}) is
equivalent to (\ref{gaugeidentity5}) and using the first structure
equation
\bea
T^a = \d e^a + \omega^a{_c} \w e^c,
\eea and (\ref{varylambda2phi}), Eq. (\ref{elimatedlambda_1_2})
becomes Eq. (\ref{gaugeidentity2}). Eq. (\ref{gaugeidentity4}) is
automatically satisfied.

I am very grateful to Robin Tucker and David Burton for helpful
discussions during my PhD study and to Faculty of Science and
Technology and Physics Department at Lancaster University for a
travel grant to attend MG11.

\vfill

\end{document}